\documentclass[aps,prl,reprint,twocolumn,superscriptaddress,floatfix,nofootinbib,longbibliography]{revtex4-2}
\usepackage{todonotes}
\usepackage{graphicx,amsmath,amsfonts,amssymb,amsthm,xr}
\usepackage{epsfig,amsmath,amssymb,color,dsfont,upgreek,physics}
\usepackage{mathrsfs}
\usepackage{mathtools}
\usepackage{bbold}
\usepackage{float}
\usepackage[caption=false]{subfig}

\usepackage[separate-uncertainty=false,per-mode=symbol,print-unity-mantissa = false]{siunitx}

\usepackage[bookmarks=true,colorlinks,linkcolor=OrangeRed,urlcolor=NavyBlue,citecolor=RoyalBlue]{hyperref}
\usepackage[dvipsnames]{xcolor}
\usepackage{orcidlink}

\usepackage{graphicx}
\usepackage{dcolumn}
\usepackage{bm}
\usepackage{color,soul}

\definecolor{mypurple}{rgb}{0.49,0.18,0.56}

\definecolor{limeGreen}{rgb}{0.55, 0.71, 0.0}

\usetikzlibrary{arrows.meta}
\usepackage{changes}
\usepackage{helvet}
\usepackage[tracking=true]{microtype}

\begin{document}
\title{Engineering Quantum Many-Body Scars through Lattice Geometry}
\author{Erick Parra Verde}
\affiliation{Department of Physics and Arnold Sommerfeld Center for Theoretical Physics (ASC), Ludwig Maximilian University of Munich, 80333 Munich, Germany}
\affiliation{Munich Center for Quantum Science and Technology (MCQST), 80799 Munich, Germany}

\author{Kevin P. Mours${}^{\orcidlink{0009-0007-3602-2198}}$}
\affiliation{Max Planck Institute of Quantum Optics, 85748 Garching, Germany}
\affiliation{Munich Center for Quantum Science and Technology (MCQST), 80799 Munich, Germany}
\affiliation{Fakult\"at f\"ur Physik, Ludwig-Maximilians-Universit\"at M\"unchen, 80799 M\"unchen, Germany}

\author{Johannes Zeiher${}^{\orcidlink{0000-0002-8466-1863}}$}
\affiliation{Max Planck Institute of Quantum Optics, 85748 Garching, Germany}
\affiliation{Munich Center for Quantum Science and Technology (MCQST), 80799 Munich, Germany}
\affiliation{Fakult\"at f\"ur Physik, Ludwig-Maximilians-Universit\"at M\"unchen, 80799 M\"unchen, Germany}

\author{Ana Hudomal${}^{\orcidlink{0000-0002-2782-2675}}$}
\email{ana.hudomal@ipb.ac.rs}
\affiliation{Institute of Physics Belgrade, University of Belgrade, 11080 Belgrade, Serbia}

\author{Jad C.~Halimeh${}^{\orcidlink{0000-0002-0659-7990}}$}
\email{jad.halimeh@lmu.de}
\affiliation{Department of Physics and Arnold Sommerfeld Center for Theoretical Physics (ASC), Ludwig Maximilian University of Munich, 80333 Munich, Germany}
\affiliation{Max Planck Institute of Quantum Optics, 85748 Garching, Germany}
\affiliation{Munich Center for Quantum Science and Technology (MCQST), 80799 Munich, Germany}
\affiliation{Department of Physics, College of Science, Kyung Hee University, Seoul 02447, Republic of Korea}

\begin{abstract}
Quantum many-body scars enable persistent non-ergodic dynamics in otherwise thermalizing systems, yet their stabilization typically relies on fine-tuned initial states or engineered Hamiltonian perturbations. Here we show that lattice geometry alone can serve as a powerful and experimentally accessible control knob for inducing and enhancing scarring. By transforming a one-dimensional chain into a quasi-one-dimensional triangle-decorated lattice, we find that the fully polarized state—normally thermalizing in the PXP model—exhibits pronounced fidelity revivals, slow entanglement growth, and strong overlap with a tower of weakly entangled eigenstates. We trace this behavior to a geometry-induced restructuring of the constrained Hilbert space, whereby the adjacency graph decomposes into hypercube subgraphs that enforce coherent population transfer and stabilize an emergent approximate $\mathrm{su}(2)$ algebra. We propose a direct implementation in programmable arrays of tweezer-trapped Rydberg atoms, where the triangle-decorated geometry can be realized using spatial light modulators and the resulting scarring dynamics probed via time-resolved measurements of excitation density. Our results establish lattice connectivity as a design principle for engineering non-ergodic dynamics in constrained quantum systems.
\end{abstract}

%\date{\today} 
\maketitle
%{\hypersetup{linkcolor=mygold}
% \tableofcontents

\textbf{\textit{Introduction.---}}
Rydberg atoms have emerged as a leading platform for quantum simulations of strongly correlated many-body systems \cite{Browaeys2020}. These setups naturally realize interacting spin models and have enabled the exploration of phenomena relevant to condensed matter and high-energy physics. A notable discovery in such experiments is quantum many-body scars (QMBS)~\cite{Bernien2017,Turner2017,Turner2018,Serbyn2021,MoudgalyaReview,ChandranReview}, a small subset of nonergodic eigenstates in an otherwise ergodic spectrum, which can significantly delay thermalization from certain initial configurations overlapping strongly with them. 
The discovery of QMBS has opened a new research direction in quantum many-body physics and their signatures have since been identified in a wide and growing range of systems, including AKLT~\cite{Moudgalya2018,Mark2020,ODea2020}, XY~\cite{Iadecola2019_2,Chattopadhyay2020,Gotta2023,Mohapatra2026}, 
bosonic~\cite{Zhao2020,bosonScars,Hummel2023,Kaneko2024,Beringer2024}, fermionic~\cite{Vafek2017,Moudgalya2020,Desaules2021,Imai2025,Pakrouski2026}, 
lattice gauge theory (LGT) models ~\cite{DesaulesSchwingerA,DesaulesSchwingerB,Aramthottil2022,osborne2024quantummanybodyscarring21d,Budde2024,Calajo2025,Hartse2025}, and various others~\cite{Shiraishi2017,Ok2019topological,Pai2019,Moudgalya2020torus,Iadecola2020,Sala2020,vanVoorden2020,McClarty2020,Lee2020,Pakrouski2020,Zhao2021,Jeyaretnam2021,Wildeboer2021,Wildeboer2022,You2022,Tang2022,Moudgalya2024,Evrard2024b,Lerose2025,Pizzi2025,Hallam2026}. Beyond the original experiments on Rydberg atom arrays~\cite{Bernien2017,Bluvstein2021}, experimental realizations of QMBS have been extended to other platforms, including superconducting qubits~\cite{Zhang2022} and Bose--Hubbard quantum simulators~\cite{Su2023}.

QMBS have been most extensively studied in the PXP model. Despite its simplicity, this model offers surprisingly rich physics. Although an equivalent hard-boson model had been formulated earlier~\cite{Fendley2024}, the PXP model gained prominence as an effective description of Rydberg atom arrays in the strongly interacting, nearest-neighbor blockade regime~\cite{Lesanovsky2012}, providing the first theoretical explanation of QMBS~\cite{Turner2017,Turner2018}.
Early studies of the PXP model have focused on the 1D case and simple initial product states known to exhibit distinctive non-ergodic dynamics due to scarring, namely the antiferromagnetic N\'eel states. 
These two states have considerable overlap with a
set of QMBS eigenstates distributed roughly equally throughout the energy spectrum. Their behavior is characterized by periodic revivals, long-lived oscillations in local observables, and a markedly slow growth of bipartite entanglement entropy.
The observed dynamics have been explained as arising from an approximate dynamical symmetry described by an su(2) algebra~\cite{Choi2019,Bull2020}. Using this approach, the revivals of the N\'eel state can be intuitively understood as an approximate precession of a large spin that oscillates between the ground and top states, in this case being the N\'eel and its symmetric partner, the anti-N\'eel state.
Other studies on the origin of QMBS have focused on semiclassical periodic orbits underlying the observed revival dynamics~\cite{Ho2019,Michailidis2020,Evrard2024,Ren2025}, in analogy to quantum scarring in single-particle systems~\cite{Heller1984},
as well as the description in terms of quasiparticle excitations~\cite{Lin2019,Iadecola2019,ChandranReview}.
Previous works have also identified regular structures in the adjacency graphs of various QMBS models~\cite{hypercubes}.

\begin{figure*}[t]
  \centering
  %\fbox{\rule{0pt}{4cm} \rule{6cm}{0pt}}
 \includegraphics[width=\linewidth]{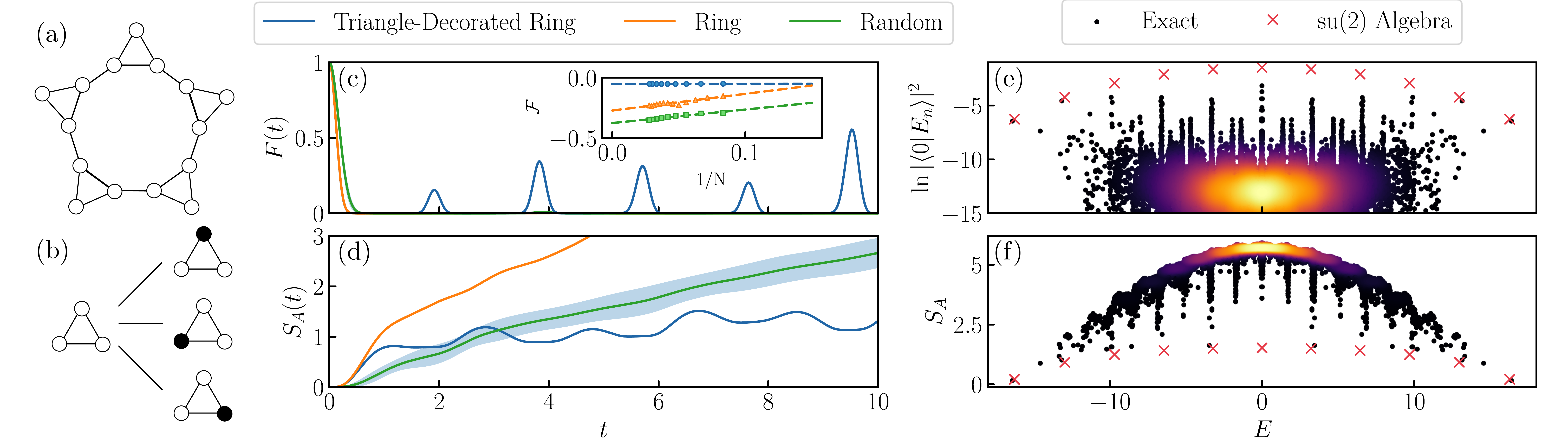}
  \caption{\textbf{Quantum many-body scarring of the polarized state in a triangle-decorated ring.} (a) Diagram of a triangle-decorated ring with $N=15$ sites, amounting to $5$ unit cells. (b) Dynamics of an isolated unit cell initialized in the polarized state. (c,d)  Quenches of the polarized state on a triangle-decorated ring (blue), the polarized state on an undecorated ring (orange) and of an ensemble of $30$ randomly chosen product states on the triangle-decorated ring (green), where the solid line represents the average and the shading represents the standard deviation. (c) Fidelity and fidelity density in the inset. (d) Bipartite entanglement entropy. Dynamical results obtained with ED on systems with $N=36$ sites. (e) Overlap between the polarized state and the eigenvectors. (f) Entanglement entropy of the eigenvectors. Colors represents the density of states in a region, with black being low density and yellow being high density. Spectrum computations were obtained with ED in the zero-momentum, inversion-symmetric sector of a triangle-decorated ring with $N=30$ sites.}  
  \label{fig:mainIdea}
\end{figure*}

The PXP model has also been studied in the context of mapping to LGT models~\cite{Surace2020}, unconventional transport properties~\cite{Ljubotina2022}, quantum criticality~\cite{YaoCriticality2022,Hudomal2025}, cellular automata~\cite{Wilkinson2020,Giudici2024}, non-stabilizerness~\cite{Smith2025}, generalizations to larger blockade radii~\cite{Kerschbaumer2025}, solitonic excitations~\cite{Kerschbaumer2025solitons}, and the effects of various perturbations~\cite{Khemani2019,Lin2020}, including disorder~\cite{MondragonShem2020}, periodic driving~\cite{Mukherjee2020,Mukherjee2020vacuum,Sugiura2021,Maskara2021,Hudomal2022Driven,Perciavalle2026}, dissipation~\cite{Jiang2025}, and a staggered potential~\cite{Desaules2023confinement}.
Potential applications of QMBS in the PXP model have also been discussed, particularly in metrology~\cite{Dooley2021,Desaules2022} and quantum machine learning~\cite{Sarkar2025}.
The extensions to 2D geometries studied so far have been mostly limited to scarring  
in bipartite lattices~\cite{Michailidis2D,Lin2020_2,Bluvstein2021,Yue2021}. The PXP model on non-bipartite lattices such as Kagome has been studied in the context of quantum spin liquids~\cite{Semeghini2021,Verresen2021,Giudici2022}.

Although it was initially believed that only a small set of simple product-state configurations exhibit quantum scarring in the PXP model, the number of known scarred states has steadily grown in recent years. The so-called polarized state, which contains no excitations and does not exhibit revivals in the pure PXP model, was shown to become scarred upon the addition of a chemical potential term~\cite{Su2023}. 
Subsequent studies have mapped out a dynamical phase diagram of the PXP model and uncovered a continuous family of QMBS states~\cite{Daniel2023,Hudomal2025}, including even the highly entangled critical state at the phase transition point.

In this work, we show that the polarized state can become scarred in the pure PXP model solely through a modification of lattice geometry from a linear chain to a quasi-1D structure with triangular decorations; see Fig.~\ref{fig:mainIdea}. Notably, this lattice is non-bipartite. This is particularly striking as it demonstrates scarring in a non-bipartite PXP lattice without N\'eel-like structure, in contrast to previously studied cases. In contrast to previous mechanisms that rely on Hamiltonian perturbations, special initial states, or engineered constraints, our results establish that lattice geometry alone suffices to induce scarring in an otherwise thermalizing initial state.
We establish the characteristic signatures of QMBS in the triangle-decorated geometry, identify the microscopic origin of this phenomenon, and propose an experimental realization of our model using Strontium Rydberg atom arrays.

\textbf{\textit{Model.---}}
An array of Rydberg atoms operating in the nearest-neighbor blockade regime is well described by the PXP model. For a general lattice geometry, the Hamiltonian takes the form

\begin{equation}
    \hat H_\text{PXP} = \Omega\sum_{i=1}^{N} \hat X_i \prod_{j:\langle i,j\rangle} \hat P_{j},
    \label{eq:1D_PXP_Hamiltonian}
\end{equation}
where $N$ is the number of sites and $\hat X_i = |0\rangle \langle 1|_i + |1\rangle \langle 0|_i$ is the standard Pauli-$x$ operator acting on site $i$. In the absence of constraints, this term would induce Rabi oscillations with frequency $\Omega$.  
A perfect nearest-neighbor Rydberg blockade is enforced by the projectors onto the ground state, $\hat P_j = |0\rangle \langle 0|_j$, which ensure that two neighboring atoms are never simultaneously dynamically excited $|\ldots11\ldots\rangle$. 
In bipartite lattices, this model exhibits QMBS when initiated in one of the antiferromagnetic states, N\'eel $|\mathbf{Z}_2\rangle=|1010\ldots10\rangle$ and anti-N\'eel  $|\mathbf{Z}'_2\rangle=|0101\ldots01\rangle$, or their superposition.
A detuning term can be included by adding $\mu \sum_{i} \hat n_i$, where $\hat n_i = \mathbb{1} - \hat{P}_i =  |1\rangle \langle 1|_i$ is the number operator on site $i$ and $\mu$ is the detuning strength. Different values of $\mu$ can induce scarring from the polarized state, $|0\rangle=|00\ldots0\rangle$, or even the ground states of the detuned Hamiltonian at different initial $\mu_0$, which are typically entangled rather than product states \cite{Su2023,Daniel2023}.
Unless specified otherwise, we assume periodic boundary conditions (PBC), corresponding to a ring geometry, and set $\Omega = 1$.

We consider a quasi-1D geometry shown in Fig.~\ref{fig:mainIdea}(a), obtained by adding an extra site, referred to as a decoration, to every other bond of the standard 1D PXP chain. This decoration renders the lattice tripartite instead of bipartite. Each triangle defines a unit cell that can host maximally one excitation, as illustrated in Fig.~\ref{fig:mainIdea}(b).
We therefore refer to this geometry as a triangle-decorated ring. 
We verify that this geometric modification keeps the model non-integrable by calculating the mean energy-level spacing ratio $\langle r \rangle = 0.53$ at $N = 36$, consistent with the Wigner--Dyson distribution (See End Matter for full level statistics, system size analysis and different decorations). 

\textbf{\textit{Scarring dynamics.---}}
At zero detuning, a quench from the polarized state in the undecorated PXP model leads to rapid thermalization consistent with ETH, in contrast to the scarred dynamics obtained from the Néel state \cite{Turner2017}. 
Here we show that modifying the lattice geometry to a triangle-decorated ring induces scarring of the polarized state, as illustrated in Fig.~\ref{fig:mainIdea}.

Figure~\ref{fig:mainIdea}(c) shows the fidelity dynamics, $F(t) = |\langle \psi(0)|\psi(t)\rangle|^2$, which measures the probability of return to the initial state $\ket{\psi(0)}$. Here we observe a stark difference between the dynamics on a ring and a triangle-decorated ring. While revivals are absent in the former case, the latter exhibits pronounced periodic revivals with an approximate period $T_\text{rev}\sim \frac{1.91}{\Omega}$. The finite-size scaling of the fidelity density at the first revival, shown in the inset and defined as $\mathcal{F} =\frac{1}{N}\ln{|\langle \psi(0)|\psi(T_\text{rev})\rangle|^2}$, further highlights this effect. The closer $\mathcal{F}$ is to zero the more robust the revival is. In the infinite system size limit $N\rightarrow\infty$, this quantity converges to a value of $-0.272$ for the undecorated ring, while for the triangle-decorated ring it approaches $-0.052$, an order of magnitude lower. 
The fidelity density is also shown to be significantly higher than what would be expected for a random state on the triangle-decorated ring, namely $-\frac{1}{N}\ln\mathcal{D}_N$, where $\mathcal{D}_N$ is the Hilbert space dimension.

Figure~\ref{fig:mainIdea}(d) illustrates the most striking difference by showing the dynamics of the bipartite von Neumann entanglement entropy, defined by $S_A(t)=-\text{Tr}\{\hat\rho_A(t)\ln[\hat\rho_A(t)]\}$, where $\hat \rho_A(t) = \text{Tr}_B\{\ketbra{\psi(t)}\}$ is the reduced density matrix of subsystem $A$, here defined as sites $n\in[0,N/2]$. The clear juxtaposition of the thermalizing dynamics on the undecorated ring, represented by a rapid growth of $S_A(t)$, against the ergodicity breaking dynamics on the triangle-decorated ring, represented by a much slower growth of $S_A(t)$, demonstrates that the change in geometry alone induces the non-ergodic behavior. Throughout Fig.~\ref{fig:mainIdea}(c,d), we also show the averaged dynamics over randomly sampled product states, a comparison to which highlights that, in this geometry, the polarized state is special. As expected from QMBS, even though quenches from the polarized state exhibit non-ergodic behavior, generic product states thermalize in accordance with ETH. 

We now turn to the spectral analysis of the PXP model on a triangle-decorated ring to confirm the observed non-ergodic dynamics indeed originates from the presence of atypical eigenstates known as QMBS. Figure~\ref{fig:mainIdea}(e) shows the overlap of the polarized state with the eigenspectrum, revealing towers of eigenstates that are approximately equally spaced in energy and carry significantly larger weight than the rest of the Hilbert space.
The red crosses correspond to an approximation which will be discussed in the following Section. These towers coincide well in energy with inverse towers of sub-entangled eigenstates, as can be seen in Fig.~\ref{fig:mainIdea}(f). Together with the dynamical signatures, the presence of such sub-entangled eigenstates with enhanced overlap supports the conclusion that this geometric modification induces QMBS on the polarized state. 

\textbf{\textit{Origins of scarring.---}}
Having established that the triangle-decorated geometry induces QMBS, we now turn to the origin of this phenomenon. The emergence of QMBS from the polarized state can be understood by analyzing the adjacency graph of the model and in terms of an emergent approximate su(2) algebra.

\begin{figure}
    \centering
    \includegraphics[width=\linewidth]{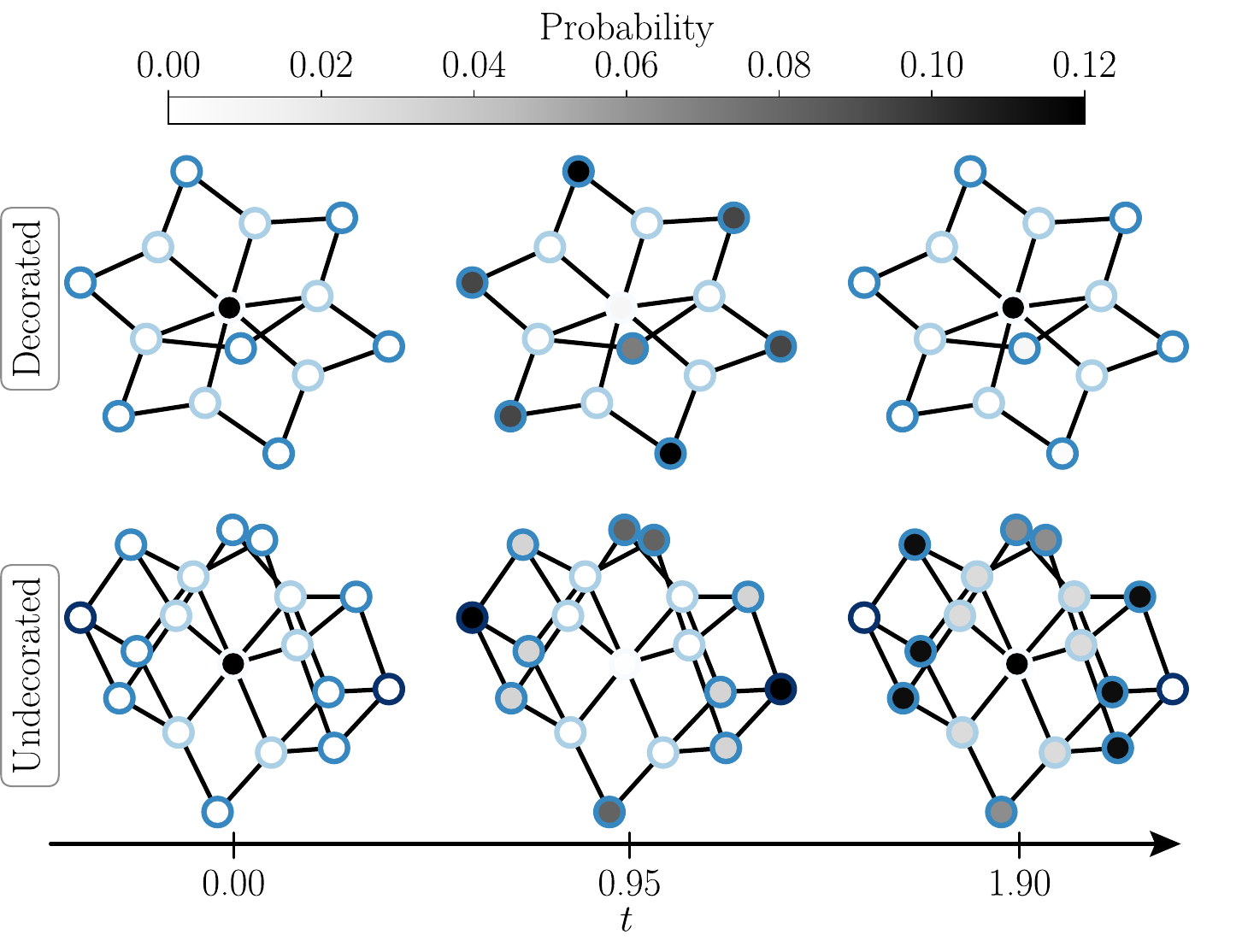}
    \caption{\textbf{Time evolution in the adjacency graph.} Top: the adjacency graph of a $N = 6$ triangle-decorated ring. Bottom: $N = 6$ undecorated ring. The outline of each node represents the density of excitations of each configuration and the inside represents the probability density of that configuration at time $t$. Three time snapshots are shown as columns. The first column shows the initial state being fully in the polarized state, represented by the central node. The second column shows the state as a superposition of highly excited configurations. The third column shows the probability density has returned to the polarized state fully for the decorated and partially for the undecorated geometry.}
    \label{fig:adjacencyFigure}
\end{figure} 

An adjacency graph in the computational basis has nodes corresponding to all Rydberg-blockade compliant product states, with edges connecting states that are related by a single action of the Hamiltonian. The adjacency graphs of both the triangle-decorated ring and the undecorated ring are shown in Fig.~\ref{fig:adjacencyFigure}. For the undecorated ring, the graph is composed of two $N/2$-dimensional hypercubes sharing a single node, which represents the polarized state, and connected by additional ``bridges'' \cite{Turner2017,hypercubes}. In contrast, the adjacency graph of the triangle-decorated ring is composed of $M_N$ $N/3$-dimensional hypercubes, glued together in various ways, but all sharing a single node corresponding to the polarized state. Here, $M_N$ denotes the number of maximally excited states, which occupy nodes antipodal to the polarized state within each hypercube. Importantly, unlike in the ring geometry, this graph contains no ``bridges'' between hypercubes, meaning that every node belongs to one of the hypercube subgraphs. From this structure, and given that an isolated hypercube supports perfect state transfer between antipodal nodes~\cite{PSTinGraphs}, we can understand the scarred dynamics starting from the polarized state as approximate state transfer between the globally shared node and the antipodal nodes across all hypercubes. Algebraically, this corresponds to coherent transfer between the polarized state and a superposition of maximally excited states. This process is illustrated in the first row of Fig.~\ref{fig:adjacencyFigure}. This hypercube connectivity enforces approximately uniform coupling between sectors of fixed excitation number, which here gives rise to the emergent approximate su(2) structure.

The equivalent process in the undecorated ring is illustrated in the second row of Fig.~\ref{fig:adjacencyFigure}. In this case, the picture of state transfer between the polarized state and antipodal nodes breaks down due to the presence of ``bridges'' that do not belong to any hypercube subgraph. Although the dynamics initially resemble transfer toward highly excited states, the absence of closed hypercube structures prevents coherent return, leading instead to spreading over many states and a rapid suppression of revivals with increasing system size, as shown in Fig.~\ref{fig:mainIdea}(c).
These ``bridges'' that spoil fidelity revivals correspond to saturated but non-maximally excited states, such as $|010010\rangle$, which admit no further excitations despite not being maximally excited. Importantly, they have a different Hamming distance to the polarized state than the maximally excited saturated states, i.e., the N\'eel states. 

In the triangle-decorated geometry, the decoration sites always allow further excitations to be added until a maximally excited state is reached, ensuring that every node belongs to a hypercube substructure. The adjacency graph also highlights the special role of the polarized state as the globally shared central node of the graph.
However, this picture alone does not fully explain the observed dynamics. While the removal of ``bridges'' ensures that all nodes belong to hypercube substructures and enables constructive interference as probability returns to the polarized state, these regular structures are not connected solely at a single node, but also share additional nodes and edges. A graph consisting of $M_N$ hypercubes joined only at one node exhibits qualitatively different dynamics. Thus, the specific manner in which the hypercubes are glued together is crucial for the emergence of QMBS. This raises the question of which features of this structure give rise to QMBS.
These observations indicate that symmetric connectivity among nodes with equal excitation number is a key ingredient for the coherent dynamics underlying the observed revivals. A complete characterization of this connectivity structure and its relation to emergent algebraic dynamics remains an interesting direction for future work.

%%%%%%%%%%%%%%%%%%%%%%%%%%%%%%%%%%%%%%%%%

We can further formalize the oscillations between the polarized and maximally excited states in terms of an emergent approximate su(2) algebra, where the polarized state and a superposition of maximally excited states are the extremal eigenstates of the operator $\hat H^z$. As discussed in more detail in the End Matter, an embedded su(2) subspace is defined by a choice of raising and lowering operators $\hat H^\pm$. It is well known that the scar states associated with the Néel state on a ring can be accurately approximated as eigenstates of a subspace generated by operators of the form~\cite{Turner2017,Choi2019,Bull2020}

\begin{equation}\label{eq:algebra_neel}
\hat H^\pm = \sum_i \Big(\hat \sigma^\pm_{2i} \prod_{j:\langle 2i,j\rangle} \hat P_{j} + \hat \sigma^\mp_{2i+1} \prod_{j:\langle 2i+1,j\rangle} \hat P_{j}\Big).
\end{equation}
For the triangle-decorated ring and the polarized state, we can instead simply define
\begin{equation}\label{eq:algebra_polarized}
    \hat H^\pm = \sum_i \hat \sigma^\pm_i \prod_{j:\langle i,j\rangle} \hat P_{j}, 
\end{equation}
which generates an $(N/3+1)$-dimensional subspace spanned by the states $|k\rangle = (\hat H^+)^k|00\ldots0\rangle$. The eigenstates in this subspace, shown as red crosses in Fig.~\ref{fig:mainIdea}(e,f), provide good approximations to the scarred eigenstates of the full system.

\begin{figure}
    \centering
    \includegraphics[width=\linewidth]{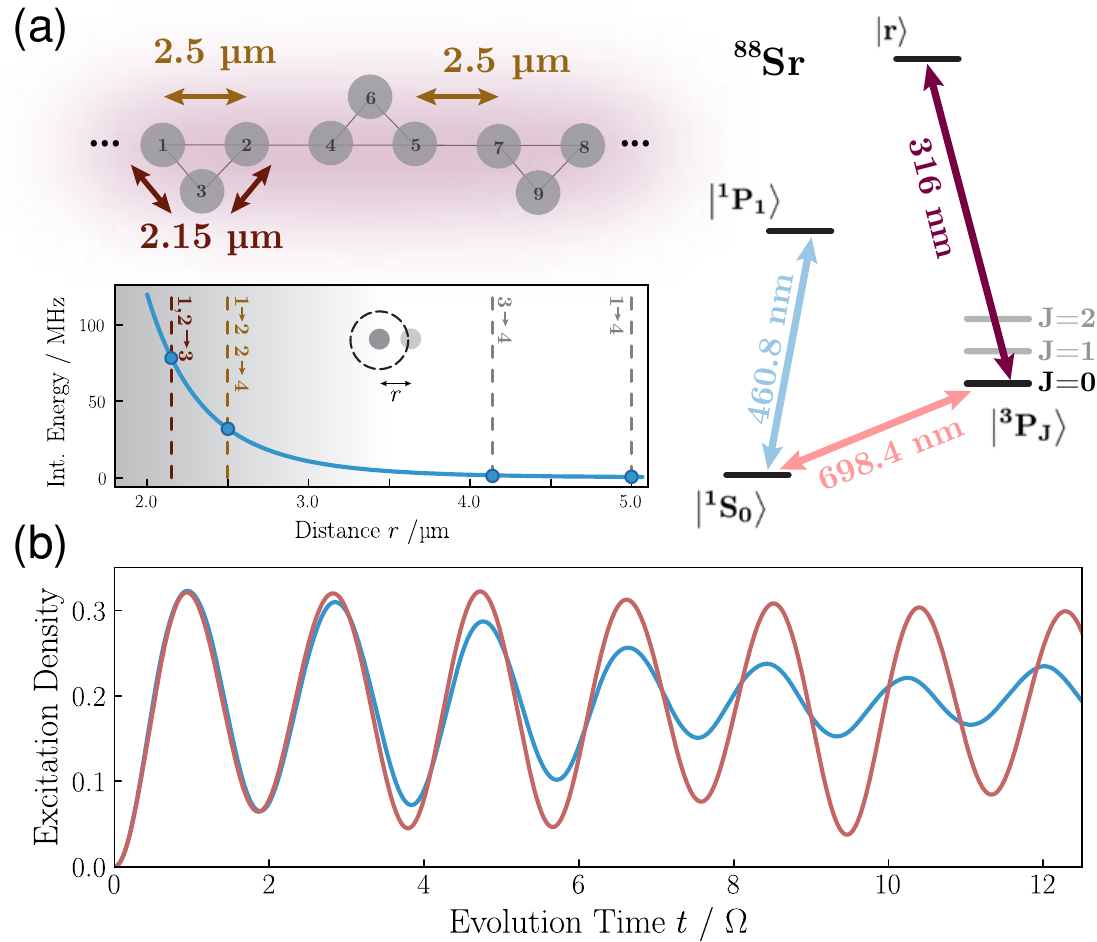}
    \caption{\textbf{Schematic of the proposed experimental setup for geometry induced scarring in a strontium-Rydberg tweezer platform} (a) Triangle-decorated chain of $^{88}$Sr atoms with nearest neighbor spacing of \SI{2.5}{\micro\meter} and decoration distance of \SI{2.15}{\micro\meter}. All atoms are simultaneously illuminated by a global UV beam (purple) with constant pulse amplitude for the duration of the evolution. Lower plot shows calculated Rydberg-Rydberg pair potential using pairinteraction \cite{pairinteraction_2017} as a function of interatomic distance. Vertical lines indicate all possible distances between atoms, where only nearest-neighbor interactions are within the blockade radius of \SI{3.5}{\micro\meter}.(b) Simulated excitation density evolution comparing the Rydberg model (blue) with the PXP model (red), illustrating the applicability for visible scarring dynamics with the shown parameters. 
    }
    \label{fig:experiment}
\end{figure}

\textbf{\textit{Experimental proposal.---}}
While the scarring dynamics is directly experimentally accessible in generic Rydberg quantum simulators~\cite{Browaeys2020}, for concreteness, in the following we focus on a specific implementation in arrays of tweezer-trapped neutral $^{88}$Sr atoms ~\cite{Norcia_2018, Cooper_2018, Madjarov_2020, Scholl_2023,Cao_2024,Shaw_2024,Sun_2026,Tao_2026}.
The required decorated lattices or chain structures can be routinely programmed with a spatial light modulator and filled deterministically with single atoms via rearrangement of a stochastically filled array.
The interatomic spacing $d$ in the chain can be chosen such that nearest neighbors lie within the Rydberg blockade radius $R_b$, approximately enforcing the constraint underlying the PXP model.
For excitation to the $n=47$ Rydberg state with Rabi frequency $\Omega/2\pi = \SI{4.3}{\mega\hertz}$, the blockade radius is estimated to be $R_b = \SI{3.5}{\micro\meter}$.
Choosing $d = \SI{2.5}{\micro\meter}$, the nearest-neighbor interaction energy $V_{NN}/2\pi = \SI{32.3}{\mega\hertz}$ exceeds the Rabi frequency, $V_{NN}\gg\Omega$, ensuring the validity of the blockade constraint.
We have numerically verified that this parameter regime maximizes the overlap between the dynamics generated by the PXP model and the full Rydberg Hamiltonian including van-der-Waals interactions  [Fig.~\ref{fig:experiment}(b)], with the required atomic geometry shown in Fig.~\ref{fig:experiment}(a).\\
To probe the scarring dynamics, the system is initialized by optically shelving all atoms into the clock state $\ket{^3\text{P}_0}$, preparing the  initial state $\ket{0}^{\otimes{N}}$ required for the protocol.
Coherent dynamics can then be driven by a global laser resonant with the $\ket{^3\text{P}_0}\rightarrow\ket{\text{r}}$ transition.
The lifetime of the $n=47$ Rydberg state is expected to be \SI{45}{\micro\second}, allowing coherent evolution over many Rabi oscillation cycles with a period of \SI{240}{\nano\second}.
The expected scarring dynamics is revealed by the excitation density evolving periodically with time.
The excitation density can be measured with high fidelity by detecting the atomic state achieved by auto-ionizing the Rydberg state $\ket{\text{r}}$ and subsequently detecting the remaining atoms with fluorescence imaging~\cite{Madjarov_2020}.

\textbf{\textit{Summary and outlook.---}}
We have shown that quantum many-body scarring can be induced in the PXP model solely through a modification of lattice geometry. By introducing a triangle-decorated structure, the fully polarized state—otherwise thermalizing—develops robust fidelity revivals, suppressed entanglement growth, and strong overlap with a tower of weakly entangled eigenstates. These signatures establish that the system exhibits genuine non-ergodic dynamics without requiring any perturbations to the Hamiltonian or special preparation of the initial state. We have traced this behavior to a geometry-induced restructuring of the constrained Hilbert space. In the decorated lattice, the adjacency graph organizes into hypercube subgraphs that share a common central node corresponding to the polarized state, while eliminating the “bridges” responsible for rapid delocalization in the standard chain. This connectivity enforces coherent population transfer between low- and high-excitation sectors and stabilizes an emergent approximate $\mathrm{su}(2)$ algebra that captures the observed dynamics.

Our results identify lattice connectivity as a minimal and experimentally accessible mechanism for stabilizing non-ergodic dynamics in constrained quantum systems. This perspective opens several directions for future work. A key question is to what extent geometry alone can be used to systematically engineer scarred dynamics in higher-dimensional or more general constrained models, including lattice gauge theories. It will also be important to develop a more complete theoretical understanding of the relation between Hilbert-space connectivity, symmetry, and emergent algebraic structures. Finally, the proposed implementation in programmable Rydberg atom arrays provides a direct route to experimentally probe geometry-induced scarring and explore its robustness in realistic settings.

\footnotesize
\medskip

\textit{Note added.---}
During completion of this work, we became aware of a complementary study that develops a graph-theoretic framework for systematically constructing quantum many-body scars in Rydberg atom arrays across arbitrary lattice geometries, identifying distinct scarring mechanisms based on lattice partitions \cite{Desaules2026}.

\medskip
\textbf{\textit{Acknowledgments.---}}
The authors acknowledge funding by the Deutsche Forschungsgemeinschaft (DFG, German Research Foundation) under Germany’s Excellence Strategy – EXC-2111 – 390814868. E.P.V.~and J.C.H.~acknowledge funding by the Max Planck Society and the European Research Council (ERC) under the European Union’s Horizon Europe research and innovation program (Grant Agreement No.~101165667)—ERC Starting Grant QuSiGauge. Views and opinions expressed are, however, those of the author(s) only and do not necessarily reflect those of the European Union or the European Research Council Executive Agency. Neither the European Union nor the granting authority can be held responsible for them. This work is part of the Quantum Computing for High-Energy Physics (QC4HEP) working group. We acknowledge funding from the Munich Quantum Valley initiative as part of the High-Tech Agenda Plus of the Bavarian State Government.
J.Z. acknowledges support from the BMFTR through the program ``Quantum technologies - from basic research to market" (Grant No. 13N17337).
K.M.~acknowledges support from the International Max Planck Research School of Quantum Science and Technology. A.H.~acknowledges funding provided by the Institute of Physics Belgrade, through the grant by the Ministry of Science, Technological Development, and Innovations of the Republic of Serbia.
\normalsize

\bibliographystyle{unsrt}
\bibliography{bibtex}

\appendix

\section{End Matter}

\subsection{Algebra}

An su$(2)$ algebraic description of the PXP model can be formulated by defining an operator $\hat H^+$ that satisfies $\hat H_\text{PXP} = \hat{H}^+ + \hat{H}^-$, where $\hat H^- = (\hat H^+)^\dagger$ \cite{Bull2020}. If these operators formed an exact su(2) algebra, the dynamics would correspond to a spin evolving under $\hat H_\text{PXP} = \hat H^x$, leading to perfect Rabi oscillations from the eigenstates of $\hat H^z$. 
However, both in the case of the N\'eel state on a ring in Eq.~\eqref{eq:algebra_neel} and the polarized state on a triangle-decorated lattice  in Eq.~\eqref{eq:algebra_polarized},
the algebra is only approximate, since
the commutation relations $[\hat H^z,\hat H^\pm ] = \pm \hat H^\pm$ are fulfilled up to error terms $\delta^\pm$. Nevertheless, these operators define an effective subspace whose eigenstates are remarkably close to the exact scarred eigenstates of the full Hamiltonian in the decorated geometry. Within this subspace, the polarized state and a superposition of maximally excited states play the role of the lowest- and highest-excited states of $\hat H^z$, respectively. The agreement can be seen in Fig.~\ref{fig:mainIdea}(e,f).

It is worth noting that this algebra is can also be defined for the polarized state in the undecorated geometry. However, in that case the error terms
$\delta^\pm$ 
are sufficiently large that the algebraic description fails to capture the dynamics as the evolution of the polarized state rapidly leaks out of the scarred su(2) subspace, preventing meaningful fidelity revivals.
In contrast, in the triangle-decorated geometry the algebra provides a significantly improved approximation. This enables an intuitive picture of the dynamics as an approximate precession of an effective $N/3$-spin between the polarized state and the equal superposition of maximally excited states, denoted by $|N/3+1\rangle$.  

The improved stability of the algebraic structure in the decorated geometry is instead reflected in the reduced coupling between the effective su(2) subspace and the rest of the Hilbert space. The magnitude of the error, naively quantified as the Frobenius norm $||\delta^\pm||_{\text{Fro}}$, does not decrease. In fact, it increases, indicating that the improvement cannot be attributed to a simple reduction of the error terms in the decorated geometry. Instead, we find that a more relevant quantity is the subspace variance normalized by the dimension of the su(2) representation, defined by 
\begin{equation}
    \frac{\sigma}{\mathcal{D}_{\text{su}(2)}} =  \frac{1}{\mathcal{D}_{\text{su}(2)}}\text{Tr}\left\{\hat U_\text{rep}^\dagger \hat H^2 \hat U_\text{rep} - \Big(\hat U_\text{rep}^\dagger \hat H \hat U_\text{rep} \Big)^2\right\},
\end{equation}
is a more relevant quantity, which decreases from $1.963$ to $0.508$ when moving from the ring to the triangle-decorated geometry at system size $N=24$. This measure quantifies the coupling between the su(2) subspace and the rest of the Hilbert space. Its reduction indicates that the su(2) subspace becomes more isolated and thus supports longer-lived coherent dynamics. For comparison, the corresponding value for the su(2) subspace associated with the Néel state on a ring is $0.112$.

\subsection{Larger Unit Cells}

\begin{figure}
  \centering
  \includegraphics[width=\linewidth]{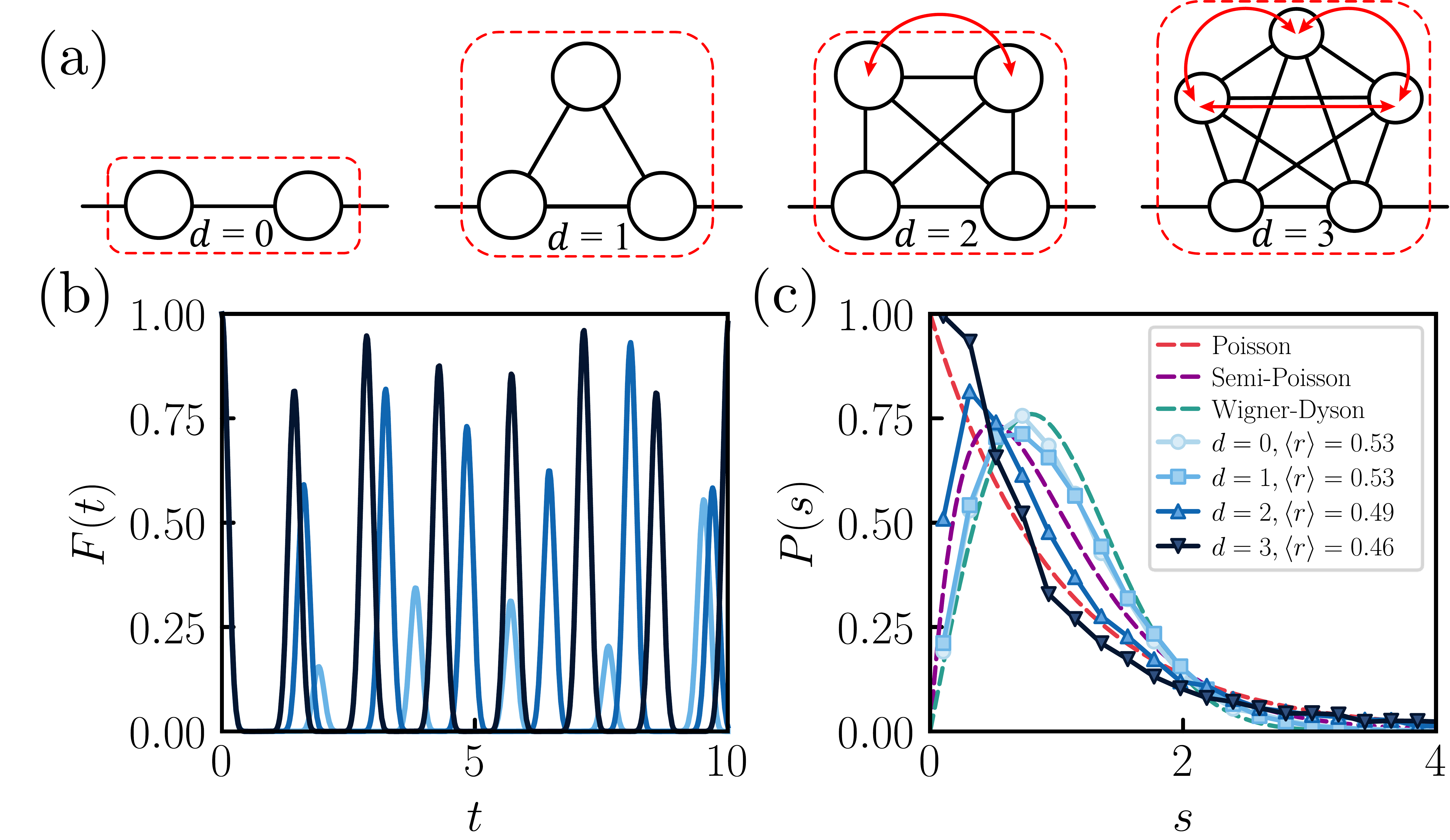 }
  \caption{\textbf{Dynamics and level statistics of bigger unit cells} (a) Diagrams of unit cells with $d$ decorations, where the permutation symmetries are highlighted in red. $d=0$ is the ring and $d = 1$ is the triangle-decorated ring. (b) Fidelity dynamics starting from the polarized state on a ring with $d$ decorations per unit cell. 
  (c) Level statistics computed in the bulk of the fully symmetric sector, i.e., the zero-momentum, inversion-symmetric, and when it applies, permutation-symmetric sector. By bulk we refer to the modes with indices $n\in [\mathcal{D}_{0+}/5,n_0-1]$ where $\mathcal{D}_{0+}$ is the Hilbert space dimension of the reduced sector and $n_0$ denotes the index of the first zero energy mode.}
  \label{fig:BiggerUnitCells}
\end{figure}

The triangle-decorated geometry can naturally be generalized to lattices with larger unit cells, where each cell remains maximally constrained internally and is connected to neighboring cells through a single site. In this case, the isolated dynamics of a unit cell initialized in the polarized state persist, with generalized oscillation between $|00\ldots0\rangle \leftrightarrow |W\rangle = \frac{1}{\sqrt{n}}\sum_{j=1}^n \sigma^+_j|00\ldots0\rangle$, for a unit cell of $n$ sites equivalent to $d = n-2$ decorations. Given that a single decoration induces QMBS of the polarized state, it is natural to ask whether larger unit cells can lead to enhanced scarring.
This question is addressed in Fig.~\ref{fig:BiggerUnitCells}.

Figure~\ref{fig:BiggerUnitCells}(a) illustrates unit cells with $d$ decorations.
Figure~\ref{fig:BiggerUnitCells}(b) shows that increasing the unit cell size enhances fidelity revivals.
However, Fig.~\ref{fig:BiggerUnitCells}(c) reveals that such models are becoming more integrable with $d$, as indicated by increasingly Poissonian level statistics. 
This behavior is to be expected, since adding multiple decorations introduces an extensive set of symmetries. In particular, unit cells with more than one decoration exhibit permutation symmetry among the decorated sites, and since this applies to every unit cell, the system acquires an extensive number of such symmetries in addition to translation and inversion.
Even after resolving these symmetries and restricting to the fully symmetric sector, as shown in Fig.~\ref{fig:BiggerUnitCells}(c), the level statistics show a clear tendency toward Poissonian behavior, indicating a crossover to integrability. Crucially, the triangle-decorated case ($d = 1$) remains fully non-integrable, demonstrating that geometry-induced scarring arises independently of integrability.

We attribute this behavior to the increasing isolation of individual unit cells. By adding more decorations, the size of each unit cell increases, while the number of connections to the other unit cells remains constant. This suggests that increasing the number of decorations effectively recovers the dynamics of an almost unconstrained two-level system evolving under $\hat X' = |0\rangle \langle W| + |W\rangle \langle 0|$. Both the extensive symmetries of the model and the reduced coupling between unit cells thus contribute to the enhanced fidelity revivals. In contrast, the absence of these permutation symmetries in the triangle-decorated model highlights this geometry as a special instance of QMBS in a fully non-integrable system with strongly interacting unit cells.

\end{document}